\documentclass[a4paper, amsfonts, amssymb, amsmath, reprint, showkeys, nofootinbib, twoside,superscriptaddress]{revtex4-1}
\usepackage[english]{babel}
\usepackage[utf8]{inputenc}
\usepackage{todonotes}
\usepackage{amsthm}
\usepackage{mathtools}
\usepackage{physics}
\usepackage{xcolor}
\usepackage{graphicx}
\usepackage[left=23mm,right=13mm,top=35mm,columnsep=15pt]{geometry} 
\usepackage{adjustbox}
\usepackage{placeins}
\usepackage[T1]{fontenc}
\usepackage{lipsum}
\usepackage{csquotes}
\definecolor{darkblue}{rgb}{0.0,0.0,0.4}
\definecolor{darkgreen}{rgb}{0.0,0.4,0.0}
\definecolor{darkred}{rgb}{0.6,0.0,0.0}
\usepackage[bookmarksopen]{hyperref}
\hypersetup{colorlinks,linkcolor=darkblue,citecolor=darkblue,urlcolor=darkblue}
\usepackage{color}
\begin{document}
\title{Dipolar droplets of strongly interacting molecules}

\author{Tim Langen}
\affiliation{5. Physikalisches Institut and Center for Integrated Quantum Science and Technology, Universität Stuttgart, Pfaffenwaldring 57, 70569 Stuttgart, Germany}
\affiliation{Vienna Center for Quantum Science and Technology,
Atominstitut, TU Wien, Stadionallee 2, 1020 Vienna, Austria}
\email{ferran.mazzanti@upc.edu}
\email{tim.langen@tuwien.ac.at}

\author{Jordi Boronat}
\affiliation{Departament de Física, Universitat Politecnica de Catalunya, Campus Nord B4-B5, 08034 Barcelona, Spain}

\author{Juan Sánchez-Baena} 
\affiliation{Departament de Física, Universitat Politecnica de Catalunya, Campus Nord B4-B5, 08034 Barcelona, Spain}

\author{Raúl Bombín}
\affiliation{Institut des Sciences Moléculaires (ISM), Université de Bordeaux, 351 Cours de la Libération, 33405 Talence, France}

\author{Tijs Karman}
\affiliation{Institute for Molecules and Materials, Radboud University, Nijmegen, The Netherlands}

\author{Ferran Mazzanti}
\affiliation{Departament de Física, Universitat Politecnica de Catalunya, Campus Nord B4-B5, 08034 Barcelona, Spain}

\date{\today}

\begin{abstract}

We simulate a molecular Bose-Einstein condensate in the strongly dipolar regime, observing the existence of 
self-bound droplets, as well as their splitting into multiple droplets by confinement-induced 
frustration. Our quantum Monte Carlo approach goes beyond the limits of the established 
effective mean-field theories for dipolar quantum gases, revealing small droplets produced 
by strong dipolar interactions outside known stable regimes. The simulations include realistic 
molecular interactions and therefore have direct relevance for current and future experiments.
\end{abstract}

\maketitle

\textit{Introduction.} Ultracold dipolar Bose-Einstein condensates (BECs) are known to exhibit many exotic quantum 
phases~\cite{Boettcher2021,Chomaz2023}, ranging from self-bound droplets~\cite{Schmitt2016,Chomaz2016} to droplet 
supersolids~\cite{Boettcher2019,Tanzi2019,Chomaz2019} and exotic supersolid patterns~\cite{Zhang2021,Hertkorn2021,Schmidt2022}. 
A key role in the formation, stability, and dynamics of these phases is played by repulsive quantum fluctuations, 
which can stabilize the dipolar systems against a mean-field collapse~\cite{Lima2011,FerrierBarbut2016}. 
In the theoretical modelling, these fluctuations are usually taken into account using an effective mean-field 
description based on the extended Gross-Pitaevskii equation  (eGPE)~\cite{Wachtler2016,Baillie2016,Saito2016}.

So far, the experimental study of dipolar gases has mostly been based on magnetic BECs, formed by weakly dipolar 
lanthanide atoms~\cite{Chomaz2023}. In these cases, the effective description yields --- except for exceptional cases~\cite{Boettcher2019b,Petter2019,Bombin2024} --- a good description of the physics observed. However, 
with the recent dramatic advances~\cite{Langen2024} in preparing strongly dipolar molecular gases close to or in the quantum 
regime~\cite{DeMarco2019,Schindewolf2022,Duda2023,Cao2023,Bigagli2023,Park2023,Lin2023,Bigagli2023BEC} 
it can be anticipated that this description will soon reach its limits~\cite{Schmidt2022,Jin2024}. 
Moreover, on a more fundamental level, it is expected that strongly interacting molecular gases will share important similarities with solid 
helium~\cite{Schmidt2022}, where vacancy-induced forms of supersolidity have been investigated for decades~\cite{Balibar2010}.
It is thus essential to explore the limits of the current effective mean-field approaches and extend the theoretical description 
of dipolar quantum matter into the strongly interacting limit. 

Here, we use quantum Monte Carlo simulations to investigate the existence of dipolar droplets in a molecular BEC. 
Using realistic interaction potentials for microwave-shielded molecules, we observe that self-bound droplets persist 
in the strongly interacting regime --- well outside the regime of validity of mean-field descriptions --- for realistic 
interaction parameters achievable in experiments. Moreover, we observe the formation of metastable droplet assemblies 
upon confinement-induced frustration, outlining a path towards molecular supersolids. Taken together, this highlights 
the prospects of molecular BECs to explore novel regimes of dipolar quantum matter.  

\begin{figure}[tb]
    \centering
    \includegraphics[width=0.46\textwidth]{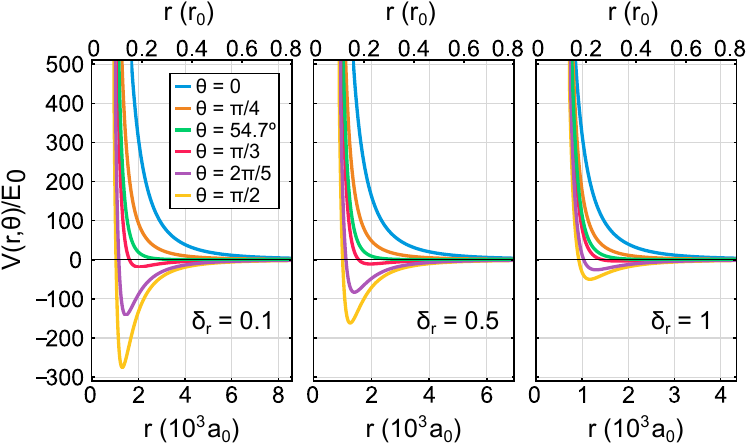}
    \caption{Molecular interaction potentials for single frequency microwave dressing of NaK molecules with a Rabi frequency of $\Omega/\hbar=2\pi\times 10\,$MHz and varying relative detuning $\delta_r$~\cite{Deng_2023}. The dressing establishes a shielding core located at around $1000\,a_0$, which prevents molecular two-body losses. Here, $a_0$ denotes the Bohr radius and $r_0$ the dipolar length, as defined in the text. By changing the dressing parameters, the potentials can be tuned and various interaction parameters can be realized, potentially resulting in different many-body states of the ultracold molecular gas.}
    \label{fig_Potential}
\end{figure}

\begin{table*}[tb]
\begin{center}
\tabcolsep = 3mm
\begin{tabular}{|l|c|ccccc|c|}
\hline
~&$\delta_r$ &   $C_6\,(10^{-3})$    & $d_{\rm eff}$ (D) &  $a_s\,(a_0)$  & $a_{\rm dd}\,(a_0) $ & $\epsilon_{\rm dd}$& $N_{\rm crit}$  \\ \hline
 (I) &  0.10    &  0.9080 & 0.7813 &  -9558.3  & 3579.9 &   -0.375 & 7    \\
  (II) & 0.50    &  1.5471 & 0.7023 &  -3644.6  & 2892.5 & -0.794 &  12    \\
   (III) & 0.75    &  2.7026& 0.6282&  -1666.1  &  2314.0 & -1.389 &  31    \\
   (IV) &1.00    &  5.0097 & 0.5552 &  -596.6  & 1807.8 & -3.030& 135    \\
   (V) & 1.25    &  9.3089 &0.4905 &  -110.1 &  1411.0 & -12.820&400    \\
\hline
\end{tabular}
\caption{Summary of the molecular interaction parameters considered in this work. We use NaK molecules and a fixed Rabi frequency of $\Omega/\hbar = 2\pi\times 10\,$MHz. By changing the relative detuning $\delta_r=|\delta|/\Omega$, the $C_6$ coefficient of the interaction, effective dipole moment $d_{\rm eff}$, $s$-wave scattering length $a_s$ and mean-field dipolar length $a_{\rm dd}$, and dipolar parameter $\epsilon_{\rm dd}=a_{\rm dd}/a_s$ can be tuned in experiments. The critical particle number $N_{\rm crit}$ is determined through our Monte Carlo simulations and marks the transition between self-bound droplets and regular gases. Notice the critical value of $C_6$ required for a two-body bound state to appear is 
$C_6^{\rm crit} \sim 0.531\times 10^{-3}$. For $C_6 > C_6^{\rm crit}$ no two-body bound state is formed, while 
droplets still exist.}
\end{center}
\label{table_drc6asNc}
\end{table*}

\textit{Molecular interactions.} Ultracold molecular gases are known to be strongly affected by two- and three-body losses at ultracold 
temperatures. These losses can, for example, occur through chemical reactions that are allowed for some species~\cite{Zuchowski2010}, or 
through the formation of short-lived complexes that can be lost from the trapping potential confining the molecules~\cite{Bause2023}. 
In contrast to ultracold atoms it is thus imperative to shield molecules from these losses at short range. A very versatile method to 
achieve this is microwave dressing~\cite{Lassabliere2018,Karman2018,Karman2019,Karman2020,Anderegg2021}, which has been used to realize 
both degenerate Fermi gases~\cite{Schindewolf2022} and, very recently, also a BEC of dipolar molecules~\cite{Bigagli2023BEC}. 
An added benefit of such shielding is the tunability of molecular interactions through a simple change of microwave power and frequency. 
Moreover, as the resulting interaction potentials are formed from microwave-dressed rotational states of the molecules, they are known 
with high precision and thus ideally suited for numerical simulations, such as the ones presented here. This is in stark contrast 
to magnetic atoms~\cite{Chomaz2023}, where the inter-atomic scattering is chaotic and the shape of the interaction potential is 
thus highly uncertain~\cite{Boettcher2019b}. 

In this Letter, we use a recently proposed analytical approximation to model the microwave-dressed molecular interactions~\cite{Deng_2023}. 
This approximation has been shown to agree very well with the full molecular interaction potentials over the parameter range relevant in 
typical experiments. We note that our simulations can easily be adapted to describe more complex shielding potentials such as, for example, 
the recently introduced double microwave dressing~\cite{Bigagli2023BEC}, DC shielding~\cite{Matsuda2020}, or combinations of MW and DC 
fields~\cite{Schmidt2022,Gorshkov2008}.

Expressed in spherical coordinates and for circularly polarized microwaves, the interaction potential of two molecules separated 
by a distance vector ${\bf r}$ is given by 
\begin{equation}
V({\bf r}) = {C_6 \over r^6} \sin^2\theta ( 1 + \cos^2\theta ) +
{C_3 \over r^3}(3 \cos^2\theta - 1) \ ,
\label{V_Deng}
\end{equation}
with $\theta$ the declination angle.
In this expression, $C_3=d^2/[48\,\pi\epsilon_0(1+\delta_r^2)]$ and $C_6=d^4/[128\,\pi^2\epsilon_0^2\, \Omega (1+\delta_r^2)^{3/2}]$ are 
functions of the relative detuning $\delta_r=|\delta|/\Omega$. Here, $\delta$ and $\Omega$ are the detuning and Rabi frequency of the 
microwave field, respectively. Furthermore, $d$ denotes the permanent dipole moment in the molecular frame and $\epsilon_0$ is the 
vacuum permeability. The resulting effective dipole moment in the lab frame is given by $d_{\rm eff}=d/\sqrt{12(1+\delta_r^2)}$, 
and oriented in the $z$-direction, perpendicular to the microwave polarization. Notice that the dipolar contribution $\sim C_3/r^3$ 
has a reversed --- \textit{anti-dipolar} --- form~\cite{Chomaz2023,Lahaye_2009,Prasad2019}. The corresponding change in sign has a 
direct influence on the final shape of the system as we discuss below.

As a concrete example, in this work we use parameters for bosonic NaK molecules~\cite{Voges2020},  with a dipole moment of $2.72\,$D, 
mass $m=62\,$amu and rotational constant $B/\hbar=2\pi\times2.089\,$GHz. For the Rabi frequency, we use $\Omega/\hbar = 2\pi\times 10\,$MHz, 
which is well-achievable in typical experiments~\cite{Schindewolf2022}. These parameters describe an interaction potential where bound states 
are located far from threshold, which is expected to minimize losses in experiments~\cite{Bigagli2023}. Example plots of the interaction 
potential are depicted in Fig.~\ref{fig_Potential} for different declination angles $\theta$ and different interaction strengths. 
The plots highlight, in particular, the large size of the repulsive shielding core $\sim 1000\,a_0$, which is in stark contrast to 
the usual contact interaction potential used to describe ultracold atoms. 

Besides the effective dipole moment $d_{\rm eff}$, changes in interaction strength are reflected in the $s$-wave scattering length $a_s$, 
which is expected to be the only relevant scattering parameter in a dilute bosonic gas at ultracold temperatures. Relating 
this last quantity to the potential parameters requires solving the low-momentum
limit of the scattering T-matrix as in Ref.~\cite{Macia_2016}, or equivalently the 
long-range behavior of the $s$-wave component of the wave function of the $E=0$ 
two-body problem. For the given potential, these calculations are more involved than usual, due to the coupling between modes 
$l, l\pm 2$ and $l\pm 4$ coming from the purely repulsive part of the interaction. The resulting values of $a_s$, and related 
to this, dipolar parameters $\epsilon_{dd}=a_{\rm dd}/a_s$, are given in Table I for typical parameters used in this work. Here, 
$a_{\rm dd}=m\, d_{\rm eff}^2/[12 \pi\epsilon_0 \hbar^2]$ denotes the dipolar length. 

In view of the large shielding core and strong dipolar interactions of the molecules, it is an interesting question whether 
the scattering length remains the relevant low-energy parameter that fully characterizes the interactions, whether renormalization 
of its strength is required as in dysprosium atoms~\cite{Oldziejewski2016,Boettcher2019b}, or whether a new approach is necessary
 altogether~\cite{Yi2001,Ticknor2008,Wang2008}. In this work, we therefore provide both purely experimental parameters such as 
detuning and Rabi frequency, in parallel with the conventional scattering length.

\label{sec_PIGS}
\textit{Simulations. }Based on the given realistic interaction potential, 
in the following we analyze a system of $N$ indistinguishable dipolar bosonic molecules in three dimensions described by the Hamiltonian
\begin{equation}
H = -{\hbar^2 \over 2m}\sum_{j=1}^N \nabla_j^2 + \sum_{i<j} V({\bf r}_{ij}) \ , 
\label{Hamiltonian}
\end{equation}
where $V({\bf r})$ is the two-body interaction potential given in Eq.~(\ref{V_Deng}). 

We use the Path Integral Ground State (PIGS) algorithm~\cite{Sarsa_2000,Rota_2010} to simulate the ground state of the system. 
It starts from a variational wave function that is propagated in imaginary time to eliminate all the components that are orthogonal 
to the true ground state, provided the propagation is carried out over a large enough imaginary time, and a good short-time approximation 
of the time-evolution operator is used. Notice that this is an {\em ab~initio} method that leads to the exact ground-state solution 
for a given Hamiltonian. Remarkably and in contrast to previous works~\cite{Boettcher2019b,Cinti2017,Kora2019,Boninsegni2021}, 
in the current case an accurate description of the Hamiltonian is known and given by the expressions above, thus allowing for a 
robust simulation of the molecular BEC's behavior. The fact that the Hamiltonian is known, makes quantum Monte Carlo methods, such as PIGS, 
an optimal choice to study this system.

In our simulations, we use dimensionless quantities expressed in terms of the 
dipolar length $r_0=m C_3/\hbar^2$ and energy $E_0=\hbar^2/mr_0^2$. Note the difference of $1/3$ in the numerical pre-factor of $r_0$ 
when comparing to the alternative dipolar length $a_{\rm dd}$, which is used in the mean-field description of dipolar gases. 
In our units convention, $C_3$ is always $1$, while the changes in the interaction parameters modify the $C_6$ parameter. 
Two-body bound states could appear when the repulsive component of the interaction is reduced below a critical value 
$C_6 \approx 0.531\times 10^{-3}$, which is, however, outside the set of parameters considered here. 

In practice, we start the simulation by setting up the system in a tightly confining trap in order to guarantee all
particles are close enough, and subsequently let the system equilibrate. Once this has been achieved, the trap is removed and the system is 
left to evolve freely. When present in the simulations, droplets are recognized by both
their energetic and structural properties: being self-bound states, the droplets total energy remains negative, while at the same time
particles stay close to each other and do not spread over the whole simulation box. These two are clear signatures of droplet formation
and allow for a good estimation of the critical atom number required for one or more droplets to form.

\begin{figure}[tb]
    \centering
    \includegraphics[width=0.48\textwidth]{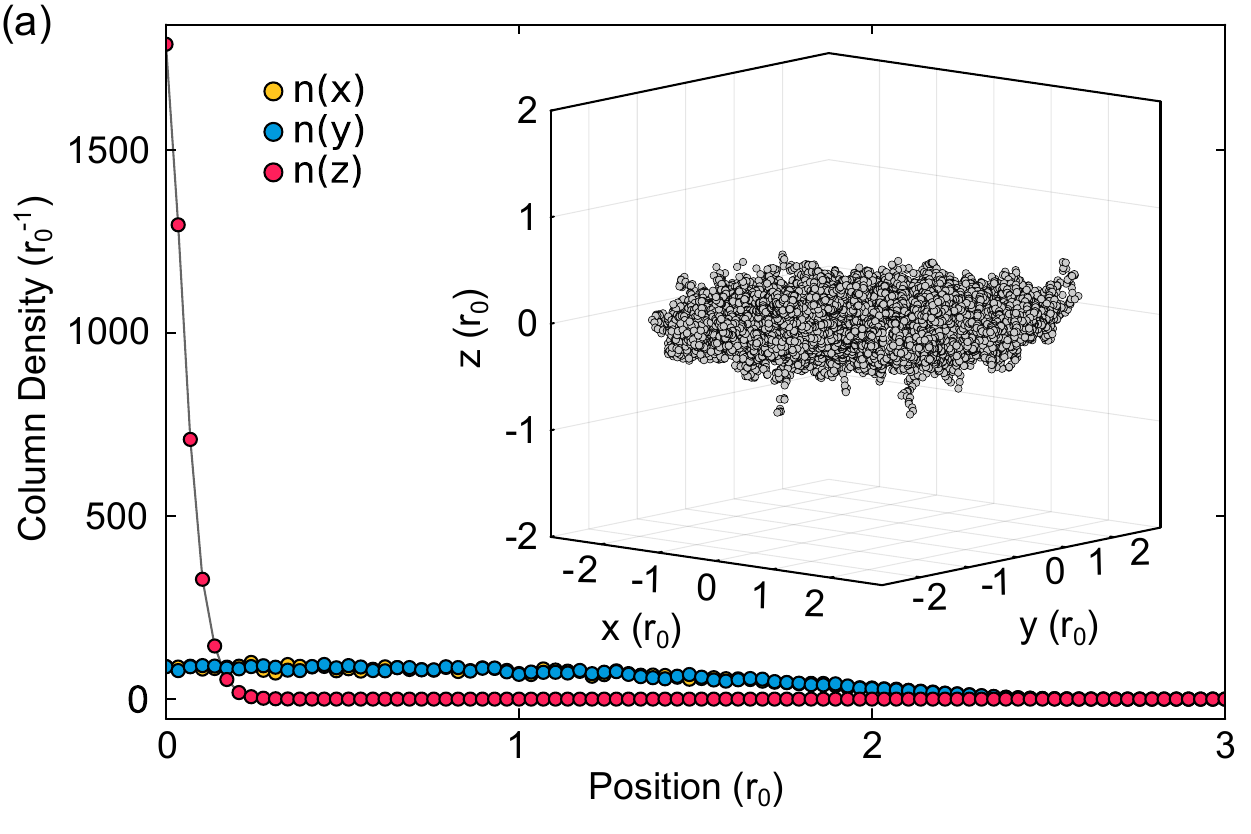}\\
    \includegraphics[width=0.49\textwidth]{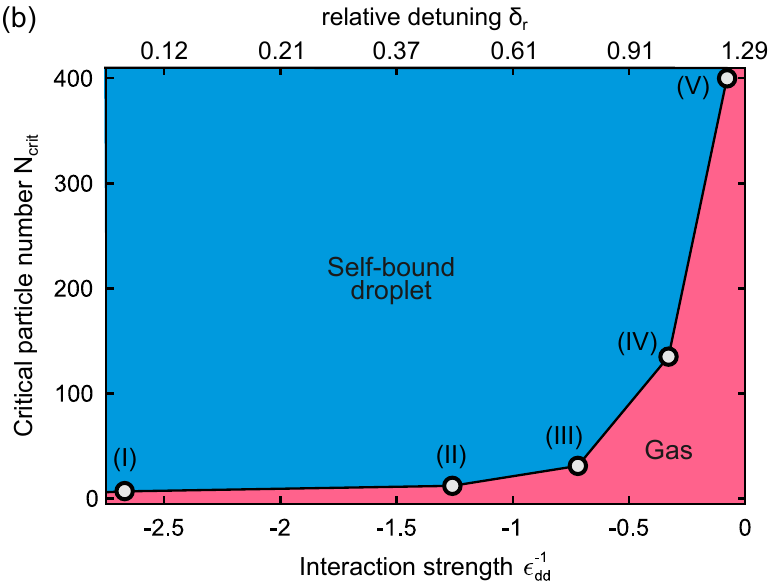}
    \caption{(a) Single self-bound dipolar droplet in the strongly interacting regime for $\delta_r=0.75$ and $N=300$ molecules. Due to the anti-dipolar 
    interactions between the molecules, the density distribution is strongly prolate. The column densities have been obtained from the average of 
    a large number of configurations, normalized to the total number of molecules. 
    The inset shows a single snapshot of a droplet. (b) Phase diagram for the transition from self-bound droplets to a normal gas phase, as a 
    function of relative detuning $\delta_r$. For reference, we also show the interaction strength parameterized by the dipolar parameter 
    $\epsilon_{\rm dd}^{-1}=a_s/a_{\rm dd}$, as it is commonly used to characterize the mean-field limit~\cite{Boettcher2019b}. 
    Parameters of the white data points (I) to (V) are summarized in Table~I.}
    \label{fig:single_drop}
\end{figure}

\textit{Results. }In Fig.~\ref{fig:single_drop}a we show example results for a simulation with $\delta_r=0.75$ and 
$N=300$ particles, leading to a stable dipolar droplet. Due to the specific form of the interaction, which has a fully 
repulsive short-ranged part combined with the sign-reversed dipolar term, the system tends to arrange 
in the plane perpendicular to the polarization of the dipoles. The observed droplet thus shows a prolate shape, 
with a radial size that is significantly larger than its minor axis. This shape is reminiscent of other, previously predicted
\textit{anti-dipolar} droplets in the mean-field regime~\cite{Prasad2019,Mukherjee2023supersolid,Kirkby2023}. Note that as the 
interaction potentials do not support a two-body bound state for the parameters considered, the droplets formed are the true 
manifestation of a many-body self-bound state~\cite{Jin2024}. 

A radial integration of the total density shows that the local gas parameter at the center of the system is
$n|a_s|^3\approx 4.36$, well beyond the reach of mean-field theory~\cite{Giorgini_1999, Mazzanti_2003, Macia_2011}. 
Remarkably, this value yields a mean inter-particle 
spacing of $n^{-1/3}\sim 50\,$nm, which is comparable to the size of the shielding core (cf. Fig.~\ref{fig_Potential}). 
This suggests that the short range details of the interaction, which are neglected in a mean-field approach, play an important 
role in the description of the droplet.  

The fact that both length scales --- the size of the shielding core and the mean interparticle spacing --- are comparable 
could also be expected to lead to three-dimensional structure formation, where the repulsively interacting molecules would 
spontaneously arrange in a crystal with the same lattice spacing~\cite{Gorshkov2008}. However, we do not find any evidence 
for this in our simulations for the parameters investigated.

Finally, the observed density $\sim6\times10^{21}\, \mathrm{m}^{-3}$ in the droplet is higher than the density in the 
molecular BEC and comparable to the densities observed in droplets of magnetic atoms~\cite{Boettcher2021}. This emphasizes 
the need to reduce three-body losses in experimental implementations to a value that is also comparable to that in atoms, 
where loss rate coefficients $L_3=1.3\times10^{-41}\, \mathrm{m}^6 \,\mathrm{s}^{-1}$ have been determined~\cite{Schmitt2016}. 

Next, we systematically map out the behavior of the system for different numbers of particles $N$ and interaction parameters. 
Our results are shown in Fig.~\ref{fig:single_drop}b. The parameters for particular values of $N_{\rm crit}$ are also summarized in Table~I. 

Similar to the behavior in magnetic atoms, where a subtle balance between dipolar interactions, contact interactions, and quantum 
fluctuations is required for stable droplets to form~\cite{Boettcher2021}, we observe a sharp transition between a parameter 
regime where self-bound droplets are prevalent, and a gaseous BEC.

Our results confirm suggestions~\cite{Schmidt2022} indicating that for increasing dipolar interaction strength the system 
supports droplets of decreasing particle number. Specifically, we find critical particle numbers as low as $N_{\rm crit}=7$ for 
$\delta_r=0.10$. The corresponding droplets are characterized by strongly attractive scattering lengths, which is in stark contrast 
to the usual mean-field description, where such parameters lead to a collapse of the system. Notably, the low particle numbers 
required are well within the range of the moderate-sized condensates that have been achieved in experiments to date~\cite{Bigagli2023BEC}.

For larger detunings of the microwave field, corresponding to weaker dipoles, the critical particle number increases, eventually 
approaching values of several hundred particles, which is similar to values previously observed in weakly-dipolar dysprosium 
gases~\cite{Boettcher2019b,Schmidt2022}. 

\begin{figure}[tb]
    \centering
\includegraphics[width=0.48\textwidth]{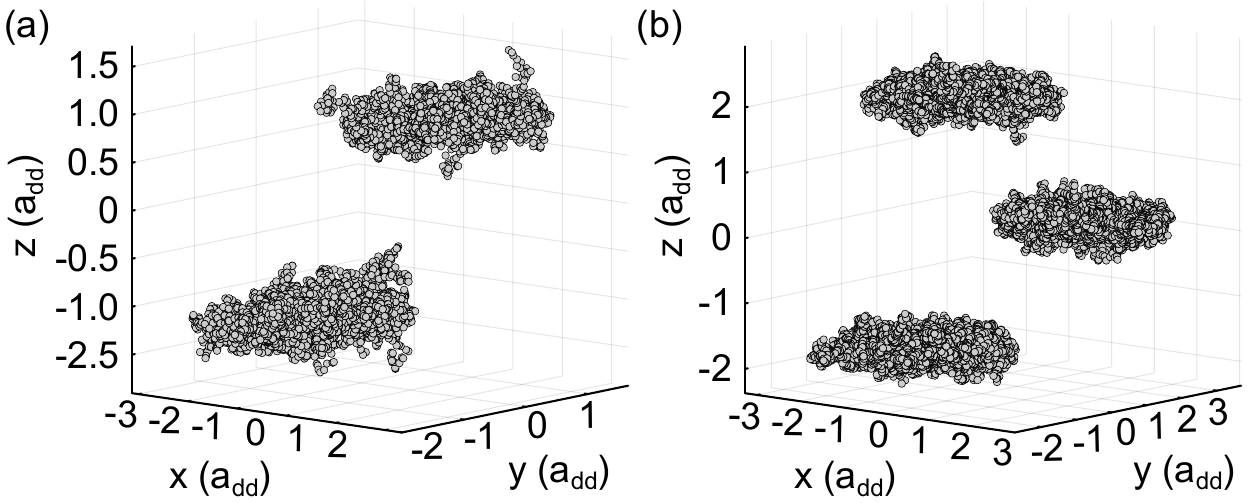}
    \caption{Snapshots of droplets formed from simulations containing (a) $N=90$ and (b) $N=300$ particles 
    for $\delta_r=0.75$, equilibrated in tight confining traps that frustrate the single droplet ground state.
    The droplets remain in a metastable state after the trap is removed.}
    \label{fig:droplets_2_and_3}
\end{figure}

Next we compare the results of the quantum Monte Carlo simulations to the predictions of the standard eGPE description, 
which requires the inclusion of a Lee-Huang-Yang (LHY) term to describe the quantum fluctuations that stabilize the droplets. 
Due to the reversed sign of the dipolar term in the interaction, however, the LHY correction term has to be rederived 
(see Appendix). The resulting critical atom numbers, obtained as the minimum number of particles
required to produce a negative energy, are far below the ones obtained from
the PIGS simulations. For $\delta_r=1.25, 1.5, 1.75$ and $2$ we find $N_{\rm crit}^{({\rm eGPE})}=9, 31, 69$ and $193$, 
respectively. The large discrepancies here are not surprising, considering the large central density
of the droplets, together with the fact that with the interaction of Eq.~(\ref{V_Deng})
and the conditions considered, the LHY correction acquires an imaginary part that can be as 
large as a 50$\%$. This again indicates that the system is well beyond a mean-field regime, even when 
the leading quantum fluctuation terms are taken into account. 

Despite the fact that in its ground state the system forms a single self-bound droplet, different metastable configurations 
can be realized by placing it in a more strongly confining trap. When the oscillator lengths
associated to the radial frequencies are smaller than the radial size of the ground state droplet, frustration occurs and 
the system splits, producing two or more nucleation centers depending on the total number of particles
and the trap frequencies employed in the initial equilibration~\cite{Wenzel2017}.
We observe a critical maximum size that each of these droplets can afford, 
favouring the splitting into more droplets when the total number of particles increases.
An example of that is shown in Fig.~\ref{fig:droplets_2_and_3}, where two snapshots of equilibrated 
configurations containing $N=90$ and $N=300$ particles are
presented. The droplets formed in this way are well isolated, with little to no exchange
of particles between them. Furthermore, despite not being the true ground state, they still have a 
very large and negative energy, thus indicating that the relaxation times required for them to evolve into
the true ground state droplet greatly exceed the time scales accessible in the simulation. These metastable 
configurations are therefore expected to be easily realized and observed in experiments, when initial tight 
confining traps are used to thermalize the system.

\textit{Conclusion. }
We have observed self-bound dipolar droplets in a realistic quantum Monte Carlo simulation 
of a molecular BEC. In particular, the results suggest that single droplets and droplet assemblies, 
which are well known from magnetic quantum gases~\cite{Boettcher2021}, exist in molecular BECs over a wide 
range of interactions parameters and strengths. In certain aspects, strongly dipolar systems can thus behave
similarly to weakly dipolar systems in the mean-field limit, as well as to other systems with competing 
interactions~\cite{Liu2008,Pupillo2010,Cinti2017,Cabrera2017}. However, the parameters and stability
regimes for the formation of many-body states can be fundamentally different, as highlighted by the observation of 
droplets even for strongly attractive short range interactions. 

We note that, within the scope of the present investigation, the system's behavior still shows a strong dependence 
on the value of the scattering length, despite being well outside the universality regime, so that other scattering 
parameters must also be relevant. More work along the lines of Refs.~\cite{Oldziejewski2016,Yi2001,Ticknor2008,Wang2008} 
is needed to explore this question in further detail. 

While the droplets realized in this work do not overlap, it will be interesting to investigate whether supersolid states also persist in the strongly interacting regime~\cite{Schmidt2022}. Certainly, the reduced number of particles per droplet makes PIGS method appropriate to study the possible BEC-supersolid transition, as it has been done recently for dysprosium system containing hundreds of atoms~\cite{Bombin2024supersolid}. Furthermore, we expect that by suitably engineering 
the confinement, a cigar shaped molecular BEC could be turned into a self-organized stack of strongly interacting layers 
with strong intra- and interlayer dipolar couplings~\cite{Mukherjee2023supersolid}, which is known to give rise to additional 
many-body phenomena~\cite{Trefzger2009,Huang2010,Du2024}.  

Our work establishes quantum Monte Carlo simulations as a tool to realistically model current and future experiments with 
ultracold molecular gases. We anticipate that systematic comparisons of such simulations, mean-field models, and experiments 
will greatly advance our understanding of dipolar quantum matter in the near future.  

~\\
\textit{Note added.---} During the preparation of this manuscript, we have become aware of the first experimental observation of droplets in a molecular BEC, including ones that are stable at negative scattering lengths~\cite{Zhang2024}.\\

We thank Jens Hertkorn, Thomas Pohl, Matteo Ciardi, and Sebastian Will for discussions, as well as Phillip Gro{\ss} and Lukas Leczek 
for a careful reading of the manuscript. T.L. acknowledges funding from the European Research Council (ERC) under the European Union’s 
Horizon 2020 research and innovation programme (Grant agreement No. 949431), from Carl Zeiss Foundation, and the Austrian Science 
Fund (FWF) 10.55776/COE1. R.B., J.S-B., F.M. and J.B. acknowledge  financial support from Ministerio de Ciencia e Innovaci\'on
MCIN/AEI/10.13039/501100011033 (Spain) under Grant No. PID2020-113565GB-C21 and from AGAUR-Generalitat de Catalunya Grant No. 2021-SGR-01411.
R.B. acknowledges funding from ADAGIO (Advanced ManufacturIng Research Fellowship Programme in the Basque – New Aquitaine Region) 
MSCA COFUND Post-Doctoral fellowship programme (Grant agreement No. 101034379).

\bibliography{biblio}

\clearpage
\appendix*
\section{The extended Gross-Pitaevskii equation for the molecular potential}

\begin{figure}[b!]
    \centering
    \includegraphics[width=0.48\textwidth]{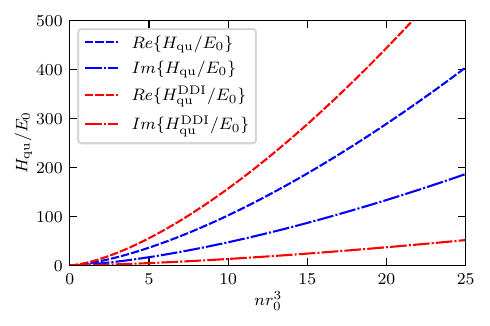}
    \caption{Real (dashed) and imaginary (dot-dashed) parts of the quantum fluctuations correction to the chemical potential $H_{\rm qu}$ for the potential of Eq.~\ref{eq_ap_pseudo} (blue) and the DDI pseudopotential (red) as a function of the homogeneous density $n$. The parameters are $a/r_0=0.11$ ($\delta_r=1.75$)}
    \label{fig_ap_1}
\end{figure}

The minimization of the energy functional with respect to the single-particle wave function within a mean field approximation yields the non-linear wave equation
\begin{widetext}
\begin{align}\label{eq_ap_0}
 &\mu \psi({\bf r}) = \left(
 -\frac{\hbar^2\nabla^2}{2m} + \!\!\int \!{\rm d}{\bf r}^\prime V_{\rm GPE}({\bf r}-{\bf r}') \abs{\psi({\bf r}')}^2 + 
 H_{\rm qu}({\bf r}) \right) \psi({\bf r}) ,
\end{align}
\end{widetext}
which determines the equilibrium state of the condensate for a given chemical potential $\mu$. The pseudopotential $V_{\rm GPE}({\bf r})$ is given by
\begin{equation}
V_{\rm GPE}({\bf r}) = g \delta({\bf r}) + {C_3 \over r^3}(3 \cos^2\theta - 1) \ ,
\label{eq_ap_pseudo}
\end{equation}
where $g=\frac{4 \pi \hbar^2 a}{m}$ is the usual contact copuling constant, with $a$ the $s$-wave scattering length.
This pseudopotential is built such that its scattering properties computed in the first-order Born approximation reproduce those of the full potential of 
Eq.~(\ref{V_Deng})~\cite{Yi2000,Yi2001}. On the other hand, the term 
\begin{align}
H_{\rm qu}({\bf r}) = \frac{32}{3} g \sqrt{\frac{a^3}{\pi}} Q_5 \left(-\frac{r_0}{3a} \right) |\psi({\bf r})|^3
\label{eq_ap_1}
\end{align}
accounts for the effect of quantum fluctuations, which provide a crucial stabilization mechanism for the otherwise collapsing system. 
The factor $3$ dividing the argument of the $Q_5$ function stems from our definition of the dipole length, which is three times larger than the usual definition $a_{\rm dd} = m C_3/(3 \hbar^2)$. Notice also that, compared to the beyond-mean-field correction for a standard dipole-dipole interaction, which we refer to as $H^{\rm DDI}_{\rm qu}({\bf r})$, the expression in Eq.~(\ref{eq_ap_1}) features an extra minus sign in the argument of the $Q_5$ function to account for the reversed sign of the dipolar term of Eq.~(\ref{V_Deng})~\cite{Schutzhold2006,Lima2011,Lima2012}. Remarkably, this reversed sign significantly changes the real and imaginary contributions of the $Q_5$ term. We illustrate this in Fig.~\ref{fig_ap_1}, where we show the real and imaginary parts of both $H_{\rm qu}$ and $H^{\rm DDI}_{\rm qu}$ computed for the homogeneous system with $a/r_0 = 0.11$ (i.e. $\delta_r = 1.75$). As it can be seen from the figure, the LHY correction for the reversed DDI shows a smaller real part, together with a significantly larger and unphysical imaginary part, which renders the application of Eq.~(\ref{eq_ap_0}) unreliable to describe the present system in the regime of parameters considered.

In order to compute the critical numbers provided in the main text, we solve Eq.~(\ref{eq_ap_0}), and subsequently 
compute numerically the Hankel transform of the wave function~\cite{Ronen2006, Mazzanti_2005}
to take advantage of the cylindrical symmetry of the problem. The critical number is determined 
by obtaining the particle number for which the energy of the ground state configuration becomes negative, 
indicating the presence of a self-bound stable solution.

\end{document}